\documentclass[runningheads]{llncs}
\usepackage{longtable}
\usepackage{xcolor}
\usepackage[T1]{fontenc}
\usepackage{graphicx}
\begin{document}
\title{AI4CAREER: Responsible AI for STEM Career Development at Scale in K-16 Education}

\author{
Sugana Chawla \and
Si Chen \and
Julia Qian \and
Gina Svarovsky \and
Alison Cheng \and
Rick Johnson \and
Nitesh V.~Chawla \and
Ronald Metoyer
}

\institute{
University of Notre Dame, Notre Dame, IN, USA
}

%
%

\maketitle              

\begin{abstract} Rapid advances in artificial intelligence (AI) are reshaping how students imagine, explore, and prepare for STEM careers across K–16 education. As AI systems increasingly influence feedback, advising, and access to opportunity information, they are becoming part of the developmental infrastructure that shapes career identity formation and readiness. Yet uncertainty remains about how AI-supported career exploration tools should be designed, governed, and evaluated at scale—particularly across developmental stages and diverse institutional contexts. This half-day workshop convenes researchers, educators, practitioners, and policymakers to examine responsible AI for STEM career development. We focus on four themes: (1) how AI reshapes definitions and assessment of STEM career readiness; (2) appropriate roles and boundaries for AI in career decision-making; (3) developmental alignment of AI supports across the K–16 continuum; and (4) equity relevant issues and design to prevent the reproduction of structural disparities. Through lightning talks, structured group activities, and cross-sector dialogue, participants will surface design tensions, articulate governance principles, and identify research gaps. The workshop aims to advance shared language and actionable frameworks for responsible, developmentally grounded AI use in STEM career learning at scale.
\end{abstract}

\section{Motivation}
Artificial intelligence (AI) is rapidly becoming embedded in the infrastructure of K–16 education. Beyond tutoring or grading support, AI systems increasingly shape how learners receive feedback, interpret performance, explore academic options, and access information about future opportunities~\cite{luckin2016intelligence,murphy2019artificial,mintz2023artificial}. Across K–16 settings—from elementary inquiry tools to secondary advising platforms and postsecondary academic support systems—AI is influencing how students make sense of their abilities and imagine possible pathways~\cite{gillani2023unpacking}. Yet these systems are often introduced without sustained scaffolding for critical engagement or developmental alignment~\cite{xie2025building}.

This shift is particularly consequential for STEM career development, which unfolds cumulatively across K–16 education. Early experiences in science and mathematics shape interest and participation, while later course-taking, specialization, mentoring, and advising translate emerging beliefs into structured educational and occupational options~\cite{kimmel2014pathways,lopez2023understanding,gandhi2016using}. However, STEM preparation remains uneven due to persistent challenges in curriculum coherence, teacher capacity, advising resources, and fragmented implementation across districts and institutions~\cite{dickman2009preparing,kimmel2014pathways}. As AI systems become embedded within instructional, exploratory, and advising infrastructures, they interact with these existing conditions and may either mitigate or amplify disparities depending on their design and deployment.

At scale, AI introduces both opportunities and risks for K–16 STEM career development. AI systems may broaden exposure to STEM pathways, personalize exploration, and surface relevant academic and career information~\cite{sun2024navigating,duan2024beyond}. At the same time, concerns arise regarding bias, inequitable access, over-personalization, and predictive labeling that may prematurely narrow students’ perceived possibilities~\cite{li2022disparities,mintz2023artificial}. Family-facing AI use further highlights tensions between autonomy and over-reliance, as well as gaps in AI literacy that complicate responsible scaffolding across developmental stages~\cite{xie2025building}. As recommendation and predictive systems increasingly generate signals about performance and potential~\cite{bahalkar2024ai}, it becomes critical to examine how these signals accumulate across K–16 transitions and shape STEM identities, aspirations, and trajectories.

Career readiness scholarship emphasizes that preparation extends beyond academic proficiency to include adaptability, exploratory behaviors, and evolving self-concepts~\cite{mishkind2014overview,camara2013defining,marciniak2022career}. If AI systems become part of the developmental infrastructure surrounding learners across K–16 education, they may influence not only short-term learning outcomes but also longer-term STEM career pathways—particularly when deployed at scale across diverse educational contexts.

This workshop centers Responsible AI for STEM Career Development at Scale in K–16 Education. We invite researchers, educators, designers, and policymakers to examine how AI systems can be designed, implemented, and governed to support cumulative STEM career development across K–16 pathways without reinforcing inequities or narrowing students’ futures.

\section{Organizers}
The organizing team brings together complementary expertise at the intersection of responsible AI, STEM education, educational assessment, advising, data science, and community-engaged learning. Collectively, we work across K–16 education, higher education, and informal STEM ecosystems, with experience spanning human-centered AI design, psychometrics, career pathway development, educator professional learning, and institutional leadership. Our work bridges research and practice, including collaborations with schools, families, industry partners, and university teaching and learning units, enabling us to examine AI not only as a technical tool but as part of broader educational and developmental infrastructure. While the current team is primarily U.S.-based, our professional networks and collaborations extend internationally. To foster broader global representation and contextual diversity, we will actively engage scholars and practitioners through established AI-in-education and STEM networks, disseminate the call across international professional societies, and connect with regional education communities to ensure participation from varied policy and cultural contexts.

\textbf{Sugana Chawla} is Associate Professor of the Practice at the Lucy Family Institute for Data \& Society at the University of Notre Dame and Data Science Education Program Director for iTREDS. With a background in Environmental Science and Education, she works at the intersection of STEM education and data science, preparing students and educators to engage meaningfully with data and emerging technologies while fostering STEM pathways.

\textbf{Si Chen} is a Postdoctoral Research Fellow at the University of Notre Dame, jointly affiliated with Notre Dame Learning and the Lucy Family Institute for Data \& Society. Her research centers on human-centered and responsible AI in K–16 STEM education. With a background in Human–Computer Interaction and experience working with students with disabilities, she uses mixed-methods and design-based research to develop AI systems that support exploration while maintaining educator oversight.

\textbf{Julia Qian} is Associate Advising Professor and Director of Advising Strategy, Assessment, and Policy in Notre Dame’s College of Engineering. Her work focuses on holistic student development and evidence-based advising. She is particularly interested in technology integration, personalized coaching, and assessment-driven policy to support student success.

\textbf{Gina Svarovsky} is Senior Executive Director for Research Engagement and Professor of the Practice at Notre Dame’s Institute for Educational Initiatives. She studies how youth develop engineering interests and skills across formal and informal STEM environments, with a focus on authentic learning experiences and pathways into engineering.

\textbf{Ying (Alison) Cheng} is the Sweeney Sweeney Family Collegiate Professor of Quantitative Psychology and Education at Notre Dame. An expert in educational assessment and psychometrics, she integrates statistical modeling, data mining, and AI to design fair and interpretable measurement systems aligned with evolving demands in quantitative literacy.

\textbf{Rick Johnson} is Associate Professor of the Practice and Managing Director of the Applied Analytics and Emerging Technology Lab at the Lucy Family Institute. He leads applied AI initiatives with industry and community partners, including K–12 STEM pathway projects and a data-focused summer internship program.

\textbf{Nitesh Chawla} is the Frank M. Freimann Professor and Lucy Family Director of Data \& AI Academic Strategy at Notre Dame. His research in AI and data science advances interdisciplinary innovation for societal impact. He is a Fellow of AAAI, ACM, AAAS, and IEEE and founder of multiple AI ventures.

\textbf{Ronald Metoyer} is Vice President and Associate Provost for Teaching and Learning and Professor of Computer Science and Engineering at Notre Dame. His work in human–computer interaction examines how data and emerging technologies can support learners and educators through responsible AI integration.
\section{Intended Audience}We aim to host approximately 25–35 participants (excluding organizers), all of whom will be co-authors of accepted workshop submissions. Participants will submit a 2–4 page work-in-progress paper or position statement using the ACM double-column template. Submissions should clearly articulate the author(s)’ perspective, ongoing work, research questions, design frameworks, empirical findings, or open challenges related to AI-supported STEM learning and career exploration. These papers will be shared with participants in advance to facilitate focused discussion, cross-site learning, and collaborative synthesis during the workshop.

As part of the submission process, authors will indicate their primary area of expertise and preferred discussion focus. This information will inform initial small-group formation. Final group composition for Activities 1 and 2 will be refined onsite to ensure balanced discussion and meaningful interdisciplinary exchange.

We aim to attract participants from a range of disciplinary communities, including Learning@Scale, Educational Data Mining, AI in Education, Human–Computer Interaction, learning sciences, STEM education, and educational policy. We particularly encourage participation from both researchers and practitioners, as a core goal of the workshop is to bridge system-level research with real-world implementation contexts in STEM learning and career exploration.

\section{Themes of Interest} This workshop is structured around four distinct but complementary themes. First, AI and the Redefinition of STEM Career Readiness focuses on theory—how AI reshapes what counts as readiness and how it should be assessed. Second, AI Design Boundaries in Career Decision-Making centers on technology governance and human interactions —what AI systems should and should not do, and where human judgment must remain central. Third, AI Across the K–16 Continuum addresses developmental alignment—how AI-supported exploration should differ by age, stage, and institutional context. Finally, AI for Equity in STEM Pathways examines structural impact—how AI can broaden participation rather than reproduce disparities.

\textbf{Redefining Career Readiness in the Age of AI} \textit{How does AI reshape how STEM career readiness is conceptualized and assessed?} We invite participants to examine how AI-mediated systems interact with developmental theories of STEM career exploration. Career readiness is multidimensional, encompassing attitudes (confidence, curiosity), knowledge (pathway awareness), and behaviors (exploration, planning)~\cite{marciniak2022career}. State and national frameworks similarly define readiness beyond content mastery~\cite{mishkind2014overview,camara2013defining}. As AI literacy becomes embedded in workforce strategies~\cite{dol2026ai}, preparation for STEM careers may increasingly include technical fluency, ethical reasoning, and human–AI collaboration. We encourage contributors to consider whether AI systems narrow readiness to measurable performance signals or expand how it is theoretically defined and longitudinally assessed.

\textbf{AI Design Opportunities, Limits and Boundaries} \textit{What roles should AI play—and not play—in STEM career decision-making?}
AI can surface pathways, prompt reflection, and provide formative feedback. However, responsible systems should not function as deterministic sorting mechanisms based on early performance signals. AI at scale must expand option spaces rather than predict “fit.” Educators and advisors remain central in interpreting recommendations and supporting identity-relevant decisions. We invite discussion on governance and evaluation: What benchmarks define effective AI-supported career coaching? Should human advising serve as a gold standard? What data infrastructures are necessary to responsibly evaluate AI systems at scale while preserving student agency?

\textbf{K–16 Differentiation and Continuum}
\textit{How should AI-supported STEM career exploration vary across developmental stages and institutions?} Across K–16 education, AI-supported exploration must align with developmental readiness. Early grades may emphasize identity expansion and exposure; middle school may scaffold exploratory behaviors; secondary and postsecondary levels may support structured decision-making tied to workforce pathways~\cite{gandhi2016using}. Because preparedness evolves iteratively~\cite{marciniak2022career}, AI systems should shift from curiosity-building to informed planning without hardening trajectories. At scale, responsible systems must remain adaptable across institutional missions, disciplinary cultures, and regional workforce contexts.

\textbf{Equity, Access, and Inclusion Pathways} \textit{How can AI systems broaden participation rather than reproduce historical inequities in STEM?} AI-driven career systems risk reproducing disparities embedded in historical data. Career development is shaped by contextual supports and access to opportunity~\cite{marciniak2022career}, and perceptions of fairness vary across student populations~\cite{li2022disparities}. We encourage discussion of bias auditing, transparent recommendation logic, diverse representation, and educator mediation~\cite{gillani2023unpacking,mintz2023artificial}. Structural barriers in STEM—including those affecting students with disabilities—must also be considered; for example, deaf and hard-of-hearing learners often face cumulative access challenges in technical terminology~\cite{chen2024towards}. Responsible AI for STEM career development should operate alongside sustained, equity-centered pipeline interventions such as \#GOALS~\cite{kuskova2025fostering}.

\section{Expected Outcomes and Contributions}

First, the organizing team will prepare a post-workshop manuscript that synthesizes insights generated across presentations and structured activities. The manuscript will distill key tensions in career guidance systems, highlight measurement and evaluation gaps, and outline priority questions for future work in large-scale educational settings. The goal is to provide a focused contribution to ongoing conversations in Learning@Scale, AI in Education, and Educational Data Mining.

Second, the workshop will generate publicly available materials. Accepted position papers will be hosted on the workshop website (with author consent), along with synthesized summaries and a structured design matrix developed during group activities. These artifacts will serve as practical reference points for researchers and practitioners examining career pathways and advising systems.

Third, we will support sustained collaboration through lightweight community infrastructure. A dedicated Slack channel will enable continued discussion and follow-up exchanges, while the workshop website will function as a stable repository for materials and updates. Together, these mechanisms will facilitate ongoing collaboration beyond the conference.

\section{Schedule} 
This 3.5-hour interactive workshop is organized around two complementary lenses: (1) understanding AI-enabled STEM career development across developmental stages and (2) proposing stakeholder-centered design solutions. During registration, participants will indicate their preferred developmental focus (middle school, high school, undergraduate) and stakeholder lens (students, families, educators/advisors, institutions/industry). Group assignments for activity 1 will be assigned and adjusted by organizers to balance expertise and ensure interdisciplinary exchange where activity 2 will be based selected by attendees. All lightning presentations are pre-submitted and grouped by the organizing team.

\textbf{0:00–0:50 (50 min) — Opening Lightning Presentations: Framing the Problem Space.}
Ten 4-minute presentations introduce key tensions in AI-supported career readiness, including shifting definitions of readiness, developmental differences, equity and access disparities, institutional constraints, and ethical boundaries.

\textbf{0:50–1:30 (40 min) — Activity 1: Developmental Lens.}
Participants break into age-based groups (middle school, high school, undergraduate). Each group analyzes stage-specific career decisions, appropriate roles for AI support, developmental guardrails, and risks of over-automation. Groups produce three opportunity areas, three risks, and three design principles, and shared with the other groups.

\textbf{1:30–1:40 (10 min) — Break.}

\textbf{1:40–2:20 (40 min) — Activity 2: Stakeholder Lens.}
Participants reorganize into stakeholder-centered groups (students, families, educators/advisors, institutions/industry). Groups examine incentive structures, authority boundaries, AI literacy requirements, and power asymmetries. Each group generates five design recommendations, one red flag to avoid, and one urgent research question, and shared with the other groups.

\textbf{2:20–3:00 (40 min) — Final Lightning Presentations: Design Solutions.}
Ten 4-minute presentations focus on AI-supported design approaches, governance strategies, and implementation models. These talks build directly on the workshop’s analytical discussions and highlight concrete pathways forward.

\textbf{3:00–3:20 (20 min) — Synthesis and Closing.}

\section{Plans for Proceedings and Advertising}

Accepted submissions will be non-archival. All accepted position papers will be published on the workshop website prior to the conference to support early engagement among participants. With authors’ permission, papers will remain publicly accessible as a resource for the broader community. Authors will be encouraged to upload their work to arXiv or similar repositories and may revise and submit their work to future peer-reviewed venues.

Organizers will promote the workshop through social media, institutional channels, and professional networks of the organizing team. The Call for Papers will be shared via platforms such as Twitter/X and LinkedIn, where organizers and collaborators will circulate the announcement within their academic and professional communities. The workshop will also be advertised through institutional websites, newsletters, and department mailing lists affiliated with the organizing team and their partner institutions.

\bibliographystyle{splncs04}
\bibliography{references}

\end{document}